\documentstyle [12pt,epsf]{article}

\input{epsf}
\begin{document}
\begin{titlepage}
\begin{center}

\vspace{-0.7in}

{\large \bf Vacuum Fluctuations of a Scalar Field \\in a Rectangular
Waveguide}\\
\vspace{.3in}{\large\em R.B.Rodrigues\footnotemark[1]}\\ 

Centro Brasileiro de Pesquisas Fisicas-CBPF\\ Rua Dr.Xavier Sigaud 150,
Rio de Janeiro, RJ,22290-180, Brazil\\
and 
\\
\vspace{.15in}{\large\em N.F.Svaiter\footnotemark[2]\footnotemark[3]}\\
Center for Theoretical Physics,\\
Laboratory for Nuclear Physics and Department of Physics,\\
Massachusetts Institute of Technology\\
Cambridge, Massachusetts 02139 USA

\subsection*{\\Abstract}
\end{center}

An analysis of one-loop vacuum fluctuations associated 
with a scalar field confined in the 
interior of a infinite waveguide of rectangular cross section 
is presented. We first consider the massless scalar field defined in 
a four-dimensional Euclidean space.  
To identify the infinities of the vacuum fluctuations 
we use a combination of dimensional and zeta function 
analytic regularization procedures. 
The infinities which occur in the one-loop vacuum fluctuations  
fall into two distinct classes: local divergences that are renormalized 
by the introduction of the usual bulk counterterms, and surface and edge 
divergences that demand countertems concentrated on the boundaries.
We present the detailed form of 
the surface and edge divergences.
Finally we discuss how to generalize our 
calculations for a confined massive scalar field defined in a 
higher dimensional Euclidean space.

\footnotetext[1]{e-mail:robson@cbpf.br}
\footnotetext[2]{e-mail:svaiter@lns.mit.edu}
\footnotetext[3]{On leave from Centro Brasileiro de Pesquisas Fisicas-CBPF}

PACS numbers:03.70+k,04.62.+v.

\end{titlepage}
\newpage \baselineskip .37in

\section{Introduction}

In 1948 
Casimir \cite{casi} predicted that uncharged, parallel, perfectly conducting 
plates should attract with a force per unit area, $F(L)\propto\frac{1}{L^{4}}$,
where $L$ is the distance between the plates. This force can be 
interpreted as the manifestation 
of the zero-point energy of the electromagnetic field
in the presence of the plates. Complete reviews of this effect can be 
found in refs. \cite{pmg} \cite{mer} \cite{livro}.

The aim of this paper is to generalize the results obtained by Fulling 
\cite{21} and others, for the renormalized vacuum fluctuations
of a massless scalar field  
calculated between two parallel flat mirrors.
Our purpose is to analyse the 
vacuum fluctuations near surfaces where the field satisfies some classical 
boundary condition, the boundary being defined by two of the coordinates 
which we assume are closed intervals.
A special case of this situation is the 
infinite waveguide of rectangular cross section.  
We are thus interested in obtaining 
the renormalized one-loop vacuum fluctuations associated 
with a  scalar field defined in the interior of a 
infinite waveguide of rectangular cross section.
We first consider a massless scalar field in a four-dimensional Euclidean 
space, and then extend our calculations to a
massive scalar field defined in a higher dimensional Euclidean space.
As stated by Deutsch and Candelas \cite{Candelas} and also Candelas \cite{refor}
the calculation of the renormalized 
one-loop vacuum fluctuations contains all the important steps that 
we need to calculate the renormalized stress-energy tensor associated with the 
scalar field.

It is well known that there are two quantities which might be 
expected to correspond to the total renormalized energy of 
quantum fields \cite{Candelas}. The first is the 
mode sum energy, defined as 
\begin{equation}
\left<E\right>_{ren}^{mode}=
\int_{0}^{\infty }d\omega \,\frac{1}{2}\omega[N(\omega)-N_{0}(%
\omega )],  \label{emode}
\end{equation}
where $\frac{1}{2}\omega $ is the zero-point energy for each mode, 
$N(\omega)d\omega $ is the number of modes with frequencies between $\omega $ and $
\omega +d\omega $ in the presence of boundaries, and $N_{0}(\omega )d\omega$
is the corresponding quantity evaluated in empty space.  
In this 
case the divergences that appear in the regularized energy are given  
by Weyl's theorem and its generalization, that relate the
asymptoptic distribution of eigenvalues of some elliptic differential
operator with geometric parameters of the surface where the fields 
satisfy some boundary condition \cite{weyl} \cite{balian}. The
second is the volume integral of the renormalized energy density $
\left<E\right>_{ren}^{vol}$ obtained by the Green's functions method 
\cite{green1} \cite{lu}
\cite{green} \cite{ben}.

For special configurations where the modes and the eigenfrequencies 
of the electromagnetic field in the presence of the boundaries 
can be found, it is possible to find the Casimir energy and the Casimir force.
For the case of cylindrical geometry, Balian and Duplantier and
also others calculated the Casimir energy of the electromagnetic field 
\cite{cyl} \cite{raad} \cite{gos} \cite{gos2}. For the important case of  
scalar or fermionic fields in the presence of a
spherical shell, Bender and Hays \cite{ben},
studying the global problem, found the renormalized zero-point energy of
these fields assuming that the fields are confined in a spherical region 
of the space. Many years before Bender and Hays, 
Boyer \cite{boy} and Davies \cite{da}
also studying the global 
problem, obtained the 
Casimir energy of an
electromagnetic field in the presence of a perfectly conducting spherical
shell. A systematic study of the spherical shell configuration was made by
Milton \cite{mi}. He calculated the zero-point energy for gluons 
and fermions, assuming that the fields are confined in the
interior of the shell. He later calculated  
the Casimir energy of
massless fermions in the presence of a spherical shell, in this case
taking into account the external modes of the fermions \cite{ex}. More 
recently, working in a generic flat d-dimensional spacetime Bender and Milton 
\cite{bm} obtained the Casimir energy associated with a massless scalar field 
taking into account the interior and exterior contributions of the 
modes in the presence of the  hypersphere.
Still studying the spherical 
configuration, Romeo \cite{romeo} investigated
the Casimir energy of a massless scalar field and for QED assuming that 
the field is confined in the interior of the spherical shell. 
Finally Bordag et.al. \cite{muitos}
have calculated the Casimir energy associated with a massive scalar field in 
the presence of a spherical shell assuming that the interior and the exterior
modes give contributions to the energy.

Although the main interest in the literature is the global 
approach, where the Casimir energy can be found and 
from which one can derive the force on the boundaries, 
the necessity of studying the local problem
has often been 
suggested.
First, the local properties of the vacuum fluctuations can 
in principle be observed by measuring the energy level 
shift of atoms interacting with an electromagnetic field \cite{weldo}
\cite{rr}. Second, the distortion of the vacuum fluctuations 
due to the presence of classical boundaries can also be measured by 
studying the spontaneous and induced emission of excited atoms in 
the presence of classical boundaries \cite{mes} \cite{lira}.
For a update discussion of QED vacuum effects, see for example \cite{qed}.
Finally, it is clear that
local results contain more information than the 
global ones. A few years ago, Actor \cite{actor} 
and also Actor and Bender considered this kind of problem  \cite{str}. 
These authors studied the use of the generalized zeta
function method \cite{ama} \cite{haw} \cite{dow} \cite{voros}
to find the one-loop effective action associated with a scalar field
defined in the interior of a infinite waveguide of rectangular cross 
section. The local problem has also been considered by other 
authors. For the case of parallel plates geometric configuration
Brown and Maclay \cite{green}, using the Green's function method 
obtained 
the renormalized vacuum expectation value of the stress-energy tensor 
associated with an electromagnetic field. 
Deutsch and Candelas \cite{Candelas} evaluated the renormalized
stress-energy tensors associated with a conformally coupled scalar field
and also with an electromagnetic field in the wedge-shaped region formed by two
plane boundaries. Recently, Brevik et al \cite{brevik} repeated these
calculations using Schwinger's source theory. Their results 
agree with those of Deutsch and Candelas. In
the spherical geometry, the local problem was also investigated by Olaussen and
Ravndal \cite{or}, who studied the vacuum fluctuations of an
electromagnetic field within a perfectly conducting spherical cavity. They
found that the vacuum expectation value of the squared electric and
magnetic fields 
diverge as one approaches the 
boundary. This result has also been obtained by 
Deutsch and Candelas \cite{Candelas},
DeWitt \cite{DeWitt}, and  
Kennedy et al \cite{sur}. 
Olaussen and Ravndal \cite{or2} and also Milton \cite{or3} generalized 
this result to the non-abelian 
gluon fields in the MIT bag model.
It has often been suggested that these surface divergences are related
to the uncertainly relation between the field and the canonical 
conjugate momentum associated with the field \cite{or} \cite{Zy} \cite{ls}.
In other local calculations, Ford \cite{idea} and Ford and Svaiter 
\cite{para} discuss the possibility 
of amplification of the vacuum fluctuations. 
These authors studied the renormalized vacuum fluctuations 
associated with a scalar and electromagnetic field 
near the focus of a parabolic 
mirror. Using the geometric optics approximation they found that
the parabolic mirror geometry can produce large 
vacuum fluctuations near the focus, 
much as what happens in the 
classical focusing effect by the parabolic mirror geometry.

In our study of the 
one-loop vacuum fluctuations  
associated with a scalar field we develop a method  
adequate
to deal with rectangular geometries. 
In this paper, using analytic regularization procedures, we first calculate 
the regularized vacuum fluctuations associated with a 
massless scalar field, 
confined in the interior of a infinite waveguide with rectangular 
cross section, in a four-dimensional Euclidean space.
We first rederive 
the well-known result that there are surface 
and edge divergences that require the introduction of surface 
\cite{Zy} \cite{di} \cite{Ne}
and edge counterterms in the renormalization procedure.
Then, we show how it is 
possible to generalize our results to a massive scalar field 
in a higher dimensional Euclidean space. 
Preliminary calculations of the renormalized vacuum 
expectation value of the stress-energy tensor in the rectangular 
waveguide were performed by Dowker and Banach \cite{banach}.
Also $\left<E\right>_{ren}^{mode}$ in rectangular 
geometries has been calculated  
in refs. \cite{caruso} \cite{pro}. A seminal paper
studying these geometries was made by Ambjorn and Wolfram 
\cite{ambb}. More recently Milton and Ng studied the Casimir effect in $(2+1)$
Maxwell-Chern-Simons electrodynamics in a rectangular domain \cite{Ng}.
Also Hacyan et al \cite{hac} and  Maclay \cite{nasa} respectively studied 
the vacuum fluctuations of the electromagnetic field and the Casimir force 
in the interior of a rectangular waveguide.

When studying interacting field theories
for translationaly invariant systems, one usually goes  
from coordinate to momentum space representation, a more convenient
framework to analyse the divergences of the theory. In this representation  
translational invariance is 
expressed by momentum conservation conditions. Because the system 
of interest for this work possesses translational invariance along 
two directions, it is more convenient to use a mixed coordinate-momentum  
representation for the Green's functions \cite{jja}. For a recent 
treatment of systems without translational invariance, 
see for example  \cite{fosco} and also \cite{jpa}.
As we discussed, 
the fundamental problem in the infinite waveguide of rectangular 
cross section  is the lack of 
translational invariance, which manifests itself by the fact that  
the Green's functions associated with the scalar field are 
expressed in terms of infinite double summations.  
Although the one-loop vacuum 
fluctuation is written in
terms of expressions involving double summations, 
a simple trigonometric identity allows us to obtain expressions with 
only one summation. The advantage is that all the calculations can then be
done analytically.  
We show that
the form of surface and edge counterterms that we have to introduce to 
renormalize the one-loop vacuum fluctuations, can be explicitly calculated
and we discuss some different 
physical arguments that support the introduction of surface countertems 
that remove these divergences.

Throughout 
the paper we use the term bulk counterterms 
to those conterterms that are not related to 
the existence of the boundaries. For instance, in the 
case of periodic boundary conditions we 
have these counterterms. As well as a  
second group that we call surface and edge counterterms that are 
related to the existence of the boundaries.

The organization of the paper is the following: In the section II we study 
the one-loop vacuum fluctuations  of a massive scalar field confined within a 
rectangular waveguide. In section
III we use an analytic regularization method to identify the infinities 
that appear in the expression of the vacuum fluctuations of the massless
field near the boundaries. 
In section IV, we show 
how it is possible to generalize our 
calculations to a massive scalar field defined in a higher dimensional 
Euclidean space. Conclusions are given in section V.
Finally in the appendix we present the formal relation 
between the one-loop vacuum 
fluctuations and the local generalized zeta function. 
In this paper we use $\hbar=c=1$.

\section{Vacuum fluctuations of a scalar field confined in a
rectangular waveguide}

In this section and in the next one we will investigate  
the one-loop vacuum fluctuations associated with a scalar field
defined in the interior of a infinite waveguide with rectangular cross 
section. First, we assume that the scalar field is defined in a 
four-dimensional Euclidean space, where the last two
coordinates are unbounded, while the first two, which we call $x_{1}$ and $
x_{2}$, lie in the interval $[0,a]$ and $[0,b]$, respectively. We assume
Dirichlet boundary conditions on the boundaries. 
The free field is defined in the region 
\begin{equation}
\Omega={{\bf x}\equiv(x_{1},x_{2},x_{3},x_{4}):0<x_{1}<a, \quad 0<x_{2}<b} 
\subset {\bf R
}^{4},
\end{equation}
with Dirichlet boundary conditions at $x_{1}=0$ and $x_{1}=a$, and also $
x_{2}=0$ and $x_{2}=b$. As stressed previously, 
the lack of translational
invariance introduces surfaces and edges divergences. 
One way to reduce the degree of these divergences is to smooth 
out the surface of the plates, for example 
by using soft, hard or  
semi-hard boundary conditions \cite{mole} \cite{alb} \cite{hard}.
The one-loop graphs
will depend on the ad-hoc model assumptions, and consequently we prefer to 
maintain the hard walls. Instead of smoothing the 
plates surfaces, another way to avoid 
surface divergences, 
discussed by Kennedy et al \cite{sur}, is to treat the boundary as a 
quantum mechanical object. This approach was used recently, by Ford and Svaiter
in the case of parallel flat plates, to
solve a long standing paradox concerning the renormalized energy of 
minimally and conformally
coupled scalar fields \cite{ls}.
As we stressed
before, we prefer, at least at the moment to keep only a hard classical
boundary conditions.
 

We would like to stress that we are 
studying first the massless four-dimensional 
case since this simpler case will give us an indication of the
behavior that may be expected in the most general case of a massive scalar 
field in a higher dimensional Euclidean space. 
A logical going is to use analytic regularization procedures to 
identify the divergent contributions that appear in the one-loop 
vacuum fluctuations. Let us start using  
dimensional regularization, working at the begining
in a d-dimensional Euclidean space. 
Assuming Dirichlet boundary conditions 
on the walls we calculate first the
two-point Schwinger function at coincident points in the 
interior of the waveguide. Then we use analytic regularization procedures 
to identify the form of the surface and edge divergences.  
In the one-loop vacuum fluctuations, $\left<\varphi^{2}(x_{1},x_{2},a,b)
\right>$, 
we will change the notation of the previous
section to $x_{1}=x$, $x_{2}=y$ (note that in our notation  
$\left<\varphi^{2}(x,y)\right>$ means the one-loop vacuum fluctuations in 
one point with cartesian coordinates $(x,y)$ of the rectangular 
cross section of the waveguide). 
Let the waveguide be 
oriented along the $z$ axis with walls at $x=0$ and $a$ and 
$y=0$ and $b$. In this
mixed representation, since we are assuming Dirichlet boundary 
conditions, the expression
for the one-loop vacuum fluctuations, 
$T_{DD}(x,y,a,b,d)=\left<\varphi^{2}(x,y)\right>$ can be written as: 
\begin{eqnarray}
T_{DD}(x,y,a,b,d)=\frac{4}{(2\pi )^{d-2}ab}\sum_{n,n^{\prime
}=1}^{\infty }\sin ^{2}( \frac{n\pi x}{a})
\sin ^{2}( \frac{n^{\prime }\pi y}{
b})  \nonumber \\
\int d^{d-2}p\frac{1}{\left( \vec{p}^{\,2}+(\frac{n\pi }{a})^{2}+(\frac{
n^{\prime }\pi }{b})^{2}+m^{2}\right) }.  \label{TadDD}
\end{eqnarray}

There are two points that we would like to stress. First 
is the fact that to perform analytic regularizations, we 
have to introduce a parameter $\mu$ with dimension of mass in order to 
have dimensionless quantities raised to a complex power. For sake 
of simplicity, we omit the $\mu$ factor in the following.
Second is the fact that the generalization for the case of Neumann 
boundary conditions is 
straightforward, although in this case infrared divergences associated with the 
$n=0$ mode will appear in the case of massless scalar 
field. To circumvent this situation 
we must have a finite Euclidean volume to regularize the model in the infrared.

Using trigonometric identities, it is convenient to write the one-loop 
vacuum fluctuation in the following way : 
\begin{equation}
T_{DD}(x,y,a,b,d)=T(a,b,d)+T(x,a,b,d)+T(y,a,b,d)+T(x,y,a,b,d),
\label{tudo}
\end{equation}
where each expression of the above equation are given by the following.  
For $T(a,b,d)$ we have 
\begin{equation}
T(a,b,d)=\frac{4}{(2\pi )^{d-2}ab}\sum_{n,n^{\prime }=1}^{\infty
}\int d^{d-2}p\frac{1}{\left( \vec{p}^{\,2}+(\frac{n\pi }{a})^{2}+
(\frac{
n^{\prime }\pi }{b})^{2}+m^{2}\right)}.  
\label{Tab}
\end{equation}
The second term in Eq.(\ref{tudo}), $T(x,a,b,d)$ is given by 
\begin{equation}
T(x,a,b,d)=-\frac{1}{2}T(a,b,d)+\frac{2}{(2\pi )^{d-2}ab}
\sum_{n,n^{\prime }=1}^{\infty }\int d^{d-2}p\frac{\cos ( \frac{2n\pi 
x}{a})
}{\left( \vec{p}^{\,2}+(\frac{n\pi }{a})^{2}+(\frac{n^{\prime }\pi }{b}
)^{2}+m^{2}\right) }.
\end{equation}
The expression for $T(y,a,b,d)$ has the same functional form of the above
equation only changing $x$ by $y$ and $a$ by $b$. Consequently we have : 
\begin{equation}
T(y,a,b,d)=-\frac{1}{2}T(a,b,d)+\frac{2}{(2\pi )^{d-2}ab}
\sum_{n,n^{\prime }=1}^{\infty }\int d^{d-2}p\frac{\cos ( \frac{2n^{\prime
}\pi y}{b})}{\left( \vec{p}^{\,2}+(\frac{n\pi }{a})^{2}+(\frac{n^{\prime
}\pi }{b})^{2}+m^{2}\right) },
\end{equation}
and finally : 
\begin{eqnarray}
T(x,y,a,b,d)=\frac{1}{(2\pi )^{d-2}ab}\sum_{n,n^{\prime }=1}^{\infty
}\int d^{d-2}p\frac{1}{\left( \vec{p}^{\,2}+(\frac{n\pi }{a})^{2}+(\frac{
n^{\prime }\pi }{b})^{2}+m^{2}\right) }  \nonumber \\
\left( 1-\cos (\frac{2n\pi x}{a})-\cos (\frac{2n^{\prime }\pi y}{b})+\cos 
(
\frac{2n\pi x}{a})\cos (\frac{2n^{\prime }\pi y}{b})\right).
\end{eqnarray}
Thus the contribution given by Eq.(8) contains 
the same contributions given by Eqs.(5),(6),(7), as well as 
a contribution that 
contains edge divergences that we will define as $N(x,y,a,b,d)$. It is given by

 
%
\begin{equation}
N(x,y,a,b,d)=\frac{1}{(2\pi )^{d-2}ab}\sum_{n,n^{\prime }=1}^{\infty
}\int d^{d-2}p\frac{\left( \cos ( \frac{2n\pi x}{a})\cos (\frac{2n^{\prime
}\pi y}{b})\right) }{\left( \vec{p}^{\,2}+(\frac{n\pi }{a})^{2}+(\frac{
n^{\prime }\pi }{b})^{2}+m^{2}\right) }.
\end{equation}
Let us study each contribution separately. Using dimensional 
regularization
on Eq.(\ref{Tab}) it is possible to write $T(a,b,d)$ in terms of the Epstein zeta 
function. Thus we have 
\begin{equation}
T(a,b,d)=\frac{4}{(2\sqrt\pi)^{d-2}ab}\Gamma (2-
\frac{d}{2})\sum_{n,n^{\prime }=1}^{\infty }\frac{1}{{\left( m^{2}+(\frac{
n\pi }{a})^{2}+(\frac{n^{\prime }\pi }{b})^{2}\right) ^{2-\frac{d}{2}}}}.
\label{acima}
\end{equation}
The contribution given by $T(a,b,d)$
is one part of the vacuum field fluctuations which does 
not depend from the distance to the boundaries and in the renormalization 
procedure will require only a  usual bulk counterterm. The form of 
the counterterm is given by the principal part of the Laurent expansion 
of Eq.(\ref{acima}) around some $d$, which must be given by the analytic 
extension of the Epstein zeta function in the complex $d$ plane. 
The structure of the divergences of the Epstein zeta function is well know 
in the literature \cite{ambb} \cite{ford} \cite{ori} \cite{kirst}. 
Since the polar structure of 
$T(a,b,d)$ can be found in the literature, to 
calculate the
analytic structure of $T(x,a,b,d)$ we will concentrate only on the 
position 
dependent divergent part given by 
$T(x,a,b,d)+\frac{1}{2}T(a,b,d)$. This expression is given by 
\begin{equation}
T(x,a,b,d)+\frac{1}{2}T(a,b,d)=\frac{2}{(2\pi )^{d-2}ab}
\sum_{n,n^{\prime }=1}^{\infty }\int d^{d-2}p\frac{\cos ( \frac{2n\pi 
x}{a})
}{\left( \vec{p}^{\,2}+(\frac{n\pi }{a})^{2}+(\frac{n^{\prime }\pi }{b}
)^{2}+m^{2}\right) }.  \label{Tedy}
\end{equation}
Although Eq.(\ref{Tedy}) is written in terms of two sums, one of the
sums can be easily done using a trigonometric expression given by 
\cite{grad}  \cite{fourier} :
\begin{equation}
\sum_{n=1}^{\infty }\frac{\cos 
nt}{n^{2}+A^{2}}=-\frac{1}{2A^{2}}+\frac{\pi 
}{2A}\frac{\cosh A(\pi -t)}{\sinh \pi A}.  \label{soma}
\end{equation}
%
Using Eq.(\ref{soma}), 
it is possible to write Eq.(\ref{Tedy}) as 
\begin{equation}
T(x,a,b,d)+\frac{1}{2}T(a,b,d)=R_{1}(a,b,d)+R_{2}(x,a,b,d)
\end{equation}
where : 
\begin{equation}
R_{1}(a,b,d)=-\frac{1}{\left( 2\pi \right) ^{d-2}ab}
\sum_{n^{\prime }=1}^{\infty }\int 
d^{d-2}p\frac{1}{(\vec{p}^{\,2}+m^{2}+(\frac{
n^{\prime }\pi}{b})^{2})}
\end{equation}
and 
\begin{equation}
R_{2}(x,a,b,d)=\frac{1}{\left( 2\pi \right) ^{d-2}b}\sum_{n^{\prime
}=1}^{\infty }\int d^{d-2}p \frac{1}{\sqrt{\vec{p}^{\,2}
+m^{2}+(\frac{n^{\prime }\pi 
}{b})^{2}}}\frac{\cosh ( (a-2x)\sqrt{\vec{p}^{\,2}
+m^{2}+(\frac{n^{\prime }\pi }{b}
)^{2}})}{\sinh (a\sqrt{\vec{p}^{\,2}+m^{2}+(\frac{n^{\prime }\pi 
}{b})^{2}})}.
\end{equation}
It is clear that to calculate the analytic structure for the case of 
the position dependent divergent part $T(y,a,b,d)$ we can use the same 
method that we use for $T(x,a,b,d)$.
Consequently the expression for $T(y,a,b,d)+\frac{1}{2}T(a,b,d)$ is : 
\begin{equation}
T(y,a,b,d)+\frac{1}{2}T(a,b,d)=I_{1}(a,b,d)+I_{2}(y,a,b,d)
\end{equation}
where : 
\begin{equation}
I_{1}(a,b,d)=-\frac{1}{\left( 2\pi \right) ^{d-2}ab}
\sum_{n=1}^{\infty }\int d^{d-2}p\frac{1}{(\vec{p}^{\,2}+m^{2}+
(\frac{n\pi }{a})^{2})}
\end{equation}
and 
\begin{equation}
I_{2}(y,a,b,d)=\frac{1}{\left( 2\pi \right) ^{d-2}a}
\sum_{n=1}^{\infty }\int d^{d-2}p\frac{1}{\sqrt{\vec{p}^{\,2}
+m^{2}+(\frac{n\pi}{a}
)^{2}}}\frac{\cosh( (b-2y)\sqrt{\vec{p}^{\,2}
+m^{2}+(\frac{n\pi }{a})^{2}})}{\sinh (
b\sqrt{\vec{p}^{\,2}+m^{2}+(\frac{n\pi }{a})^{2}})}.  
\label{I2}
\end{equation}
Using dimensional regularization in both expressions 
$I_{1}(a,b,d)$ and $R_{1}(a,b,d)$, we obtain a special Epstein-Hurwitz zeta 
function. The analytic extension of this function for general $d$ in the 
massive and massless case can be found in the literature. For the massive 
case see, for example \cite{eli}. For the massless case, 
the duplication formula for the Gamma function allows us 
to write
\begin{equation}
I_{1}(a,b,d)|_{m=0}=\frac{1}{a^{d-3}b} f_{1}(d)\zeta(4-d)\Gamma(4-d),
\end{equation}
where 
\begin{equation}
f_{1}(d)=-\frac{1}{2}
\frac{
\pi^{\frac{d-5}{2}}
}
{\Gamma(
\frac{5-d}{2})
},
\end{equation}
is an entire function of $d$. 
Since  $\zeta(z)$ can be analytically 
continued from an open connected 
set of points in the complex $z$ plane into the entire 
domain of $z$, it is easy to find the analytic structure of $I_{1}(a,b,d)|_{m=0}$
in the complex $d$ plane. 
It is clear that $I_{1}(a,b,d)$ and  $R_{1}(a,b,d)$ demand bulk counterterms.
To find the analytic structure of 
$I_{2}(y,a,b,d)$ and $R_{2}(x,a,b,d)$, let us concentrate on 
$I_{2}(y,a,b,d)$. Integrating over the
solid angle in Eq.(\ref{I2}), i.e using the 
fact that $d^{d}p=p^{d-1}dp\,d\Omega_{d}$ and $
\int \, d\Omega_{d}=\frac{2\pi^{\frac{d}{2}}}{\Gamma(\frac{d}{2})}$ 
we have :
\begin{equation}
I_{2}(y,a,b,d)=\frac{1}{a}h(d)\sum_{n=1}^{\infty }\int 
dp\,p^{d-3}\frac{1}{
\sqrt{\vec{p}^{\,2}+m^{2}+(\frac{n\pi }{a})^{2}}}\frac{\cosh ((b-2y)\sqrt{
\vec{p}^{\,2}+m^{2}+(\frac{n\pi }{a})^{2}})}{\sinh (b\sqrt{
\vec{p}^{\,2}+m^{2}+(\frac{n\pi }{
a})^{2}})}
\label{poss}
\end{equation}
where the function $h(d)$ is given by :
\begin{equation}
h(d)=\frac{2}{(2\sqrt{\pi} )^{d-2}\Gamma(\frac{d-2}{2})}.
\label{hd}
\end{equation}
Performing a change of variables $v=\left(\vec{p}^{\,2}+m^{2}+(\frac{n\pi 
}{a}
)^{2}\right) ^{\frac{1}{2}}$, and now going back to the four-dimensional 
case it is possible to write 
$I_{2}(y,a,b,d)|_{d=4}\equiv I_{2}(y,a,b)$ as :  
\begin{equation}
I_{2}(y;a,b)=\frac{1}{2\pi a}\sum_{n=1}^{\infty }\int_{\alpha}^{\infty 
}dv\,
\frac{\cosh((b-2y)v)}{\sinh bv}
\label{nota}
\end{equation}
where the lower limit of the above integral is given by  
\begin{equation}
\alpha=\sqrt{m^{2}+(\frac{n\pi}{a})^{2}}.
\label{alfa}
\end{equation}
It is important to keep in mind 
that the situation is completely different for $d \neq 4$, 
since after change of variables 
the term
$(v^{2}-m^{2}-
(\frac{n\pi }{a})^{2})^{\frac{d-4}{2}}$ it 
will appear in the integrand of Eq.(\ref{poss}). As a consequence of this fact, 
it is more difficult to perform algebraic manipulations that allow us 
to analytically regularize $R_{2}(x,a,b,d)$ and  $I_{2}(y,a,b,d)$ 
and also other expressions. 
%
%
Nevertheless, 
utilizing the same techniques used in refs. \cite{ss} \cite{claude} 
\cite{nucle},   
it is possible to perform all the calculations for a higher dimensional 
Euclidean space.  This generalization will be presented in section IV.  
Going back to Eq.(\ref{nota}) and 
using trigonometric identities, we have : 
\begin{equation}
I_{2}(y,a,b)=I_{21}(y,a)+I_{22}^{+}(y,a,b)+I_{22}^{-}(y,a,b)
\end{equation}
where :
\begin{equation}
I_{21}(y,a)=\frac{1}{4\pi ya}\sum_{n=1}^{\infty }exp \left( -2y
\sqrt{m^{2}+(\frac{n\pi }{a})^{2}}, \right),
\end{equation}
\begin{equation}
I_{22}^{+}(y,a,b)=\frac{1}{4\pi ya}\sum_{n=1}^{\infty
}\int_{b\alpha}^{\infty }dq\left( \coth q-1\right) e^{\frac{2yq}{b}},
\end{equation}
and finally 
\begin{equation}
I_{22}^{-}(y,a,b)=\frac{1}{4\pi ya}\sum_{n=1}^{\infty}
\int_{b\alpha }^{\infty }dq
\left( \coth q-1\right) e^{-\frac{2yq}{b}}.
\end{equation}
An exact expression in the massless case can be obtained from 
$I_{21}(y,a)$.
Summing the geometric series $I_{21}(y,a)|_{m=0}$ can be written as
\begin{equation}
I_{21}(y,a)|_{m=0}=\frac{1}{4\pi ay}\frac{1}{e^{\frac{2\pi y}{a}}-1}.
\end{equation}
We will use the Bernoulli polynomials and numbers 
in the next sections. Consequently, let us introduce this Laurent
expansion to study the behavior of $I_{21}(y,a)|_{m=0}$ in the neighborhood of
$y=0$. To find the principal part of the Laurent series of 
$I_{21}(y,a)|_{m=0}$  around $y=0$ let us use the Bernoulli polynomials, which 
are defined by the generating function
\begin{equation}
\frac{t\,e^{xt}}{e^{t}-1}=\sum_{n=0}^{\infty}B_{n}(x)\frac{t^{n}}{n!}, \,\, |t|<2\pi,
\label{ber}
\end{equation}
and the Bernoulli numbers $B_{n}=B_{n}(x=0)$, $(B_{0}=1,B_{1}=-\frac{1}{2}, ..)$.
Using this Laurent expansion one finds
\begin{equation}
I_{21}(y,a)|_{m=0}=\left
(\frac{B_{0}}{8\pi^{2}y^{2}}+
\frac{B_{1}}{4\pi ya}+\frac{1}{2\pi}\sum_{n=2}^{\infty}\frac{B_{n}}{n!}
(\frac{2\pi y}{a})^{n-1}\right).
\label{ber2}
\end{equation}
We note that we still have to calculate 
$N(x,y,a,b)$. Using again the trigonometric 
identity
given by Eq.(\ref{soma}), it is possible to write  
$N(x,y,a,b,d)=N_{1}(x,a,b,d)+N_{2}(x,y,a,b,d)$ where we have :
\begin{equation}
N_{1}(x,a,b,d)=-\frac{1}{2(2\pi )^{d-2}ab}
\sum_{n=1}^{\infty }\int d^{d-2}p\frac{\cos (\frac{2n\pi x}{a})}
{\left( \vec{
p}^{\,2}+m^{2}+(\frac{n\pi }{a})^{2}\right) }
\end{equation}
and
\begin{eqnarray}
N_{2}(x,y,a,b,d) &=&\frac{1}{(2\pi )^{d-1}a}
\sum_{n=1}^{\infty }\int d^{d-2}p\frac{\cos (\frac{2n\pi x}{a})}
{\sqrt{\left(\vec{
p}^{\,2}+m^{2}+(\frac{n\pi }{a})^{2}\right)}}  \nonumber \\
&&\frac{\cosh ((b-2y)\sqrt{\vec{p}^{\,2}+m^{2}+(\frac{n\pi }{a})^{2}})}{
\sinh (b\sqrt{\vec{p}^{\,2}+m^{2}+(\frac{n\pi }{a})^{2}})}.
\end{eqnarray}
Let us study the expression given by $N_{1}(x,a,b,d)$. 
It is possible to write $N_{1}(x,a,b,d)=N_{11}(a,b,d)
+N_{12}(x,a,b,d)$,
where  $N_{11}(a,b,d)$ and $N_{12}(x,a,b,d)$ are given by
\begin{equation}
N_{11}(a,b,d)=\frac{1}{4(2\pi )^{d-2}ab}\int d^{d-2}p\frac{
1}{(\vec{p}^{\,2}+m^{2})},  \label{N11}
\end{equation}
and
\begin{equation}
N_{12}(x,a,b,d)=-\frac{1}{4(2\pi )^{d-2}b}\int d^{d-2}p\frac{1}{
\sqrt{\vec{p}^{\,2}+m^{2}}}\frac{\cosh ((a-2x)\sqrt{\left( \vec{p}
^{\,2}+m^{2}\right) })}{\sinh (a\sqrt{\vec{p}^{\,2}+m^{2}})}.
\label{n12}
\end{equation}
The expression given by Eq.(\ref{N11}) can be easily calculated using
dimensional regularization and can be written as
\begin{equation}
N_{11}(a,b,d)=\frac{1}{4ab(2\sqrt{\pi})^{d-2}}\Gamma(2-\frac{d}{2})
(m^{2})^{\frac{d}{2}-2}.
\label{n11}
\end{equation}
This contribution for the vacuum fluctuations also demands bulk conterterm.
We have to deal with the expression of $
N_{12}(x,y,a,b,d)$.
Integrating over the solid angle, changing the variables and using the 
fact
that we are in $d=4$, i.e. defining 
$N_{12}(x,a,b,d)|_{d=4}\equiv N_{12}(x,a,b)$ we have
\begin{equation}
N_{12}(x,a,b)=-\frac{1}{8\pi b}\int_{m}^{\infty }dv\frac{\cosh
((a-2x)v)}{\sinh av}.
\end{equation}
Again, using trigonometric identities, we have : 
\begin{equation}
N_{12}(x,a,b)=-\frac{1}{8\pi ab}\left[ \frac{a}{2x}e^{-2xm}+\frac{1
}{2}\int_{am}^{\infty }dq\left( \coth q-1\right) 
e^{2q\frac{x}{a}}+\frac{1}{2
}\int_{am}^{\infty }dq\left( \coth q-1\right) e^{-\frac{2qx}{a}}\right].
\label{N12}
\end{equation}
We note that Eq.(\ref{N12}) 
has surface divergences when $x\rightarrow 0$ and $x\rightarrow a$. 
The structure of the divergences of $N_{12}(x,a,b)$ will be studied 
further. 

As a next step in the discussion, let us investigate the part of the 
one-loop vacuum fluctuations that contains edge divergences that is given by
$N_{2}(x,y,a,b)$. Again, integrating over the
solid angle, changing the variables and using the fact that we are in 
$d=4$,
the expression for $N_{2}(x,y,a,b)|_{d=4}\equiv N_{2}(x,y,a,b)$ is given 
by 
\begin{equation}
N_{2}(x,y,a,b)=\frac{1}{4a}\sum_{n=1}^{\infty }\cos (\frac{2n\pi x}{a}
)\int_{\alpha }^{\infty }dv\,\frac{\cosh ((b-2y)v)}{\sinh bv},
\end{equation}
where the lower limit of the above integral is given by 
$\alpha$, defined in Eq.(\ref{alfa}).
Using trigonometric identities it is possible to write $N_{2}(x,y,a,b)$ as 
\begin{equation}
N_{2}(x,y,a,b)=N_{21}(x,y,a)+N_{22}(x,y,a,b), 
\end{equation}
where : 
\begin{equation}
N_{21}(x,y,a)=-\frac{1}{8ay}\sum_{n=1}^{\infty }\cos (\frac{2n\pi x}{a}
)e^{-2y\alpha }  \label{N21}
\end{equation}
and
\begin{equation}
N_{22}(x,y,a,b)=\frac{1}{4a}\sum_{n=1}^{\infty }\cos (\frac{2n\pi x}{a}
)\int_{\alpha }^{\infty }dv\,\left( \coth (bv)-1\right) \cosh 2vy.  
\label{N22}
\end{equation}
From now we will put $m=0$. 
Using the Poisson's kernel given by 
\begin{equation}
1+2\sum_{n=1}^{\infty}r^{n}\cos n\phi= 
\frac{1-r^{2}}{1-2r\cos\phi+r^{2}}, \,\,\,  0\leq r <1,
\label{po}
\end{equation}
it is posssible to 
obtain a closed expression 
for Eq.(\ref{N21}), for points outside 
the boundaries. 
Note that $N_{21}(x,y,a,b)$ diverges for $y\rightarrow 0$ for any $x$. Let 
us study the behavior of $N_{21}(x,y,a,b)$ near the edges $x=0,y=0$ and
also $x=a,y=0$. Using the Poisson's kernel, 
the function  $N_{21}(x,y,a,b)$ in $x=0$ or $x=a$ is
given by 
\begin{equation}
N_{21}(x,y,a)|_{x=0}=\frac{1}{16ay}-
\frac{1}{16ay}\frac{e^{\frac{4\pi\,y}{a}}}{e^{\frac{4\pi\,y}{a}}+1}+
\frac{1}{16ay}\frac{1}{e^{\frac{4\pi\,y}{a}}+1}.
\label{po2}
\end{equation}
Again using the generating function of the Bernoulli numbers it is possible
to find the principal part of the Laurent expansion of 
$N_{21}(x,y,a,b)|_{x=0}$  around $y=0$, and also $N_{21}(x,y,a,b)|_{x=a}$
around $y=0$. The Laurent expansion around $y=0$ allow us to write 
$N_{21}(x,y,a,b)|_{x=0}$ as
\begin{equation}  
N_{21}(x,y,a)|_{x=0}=\frac{B_{0}}{64\pi\,y^{2}}+\frac{B_{1}}{16ay}+
O(y).
\label{po3}
\end{equation}
We note that Eq.(\ref{N21}) has two edge
divergences, one at $x=y=0$ and the other at $x=a,$ $y=0$. 
To investigate the divergences of Eq.(\ref{N22}) 
let us rewrite $N_{22}(x,y,a,b)$ as 
\begin{equation}
N_{22}(x,y,a,b)=N_{22}^{+}(x,y,a,b)+N_{22}^{-}(x,y,a,b)
\end{equation}
where 
\begin{eqnarray}
N_{22}^{+}(x,y,a,b) &=&\frac{1}{8ab}\sum_{n=1}^{\infty }\cos (\frac{2n\pi 
x}{
a})\int_{\alpha b}^{\infty }dq\,\left( \coth q-1\right) e^{\frac{2qy}{b}}
\label{N22+} \\
N_{22}^{-}(x,y,a,b) &=&\frac{1}{8ab}\sum_{n=1}^{\infty }\cos (\frac{2n\pi 
x}{
a})\int_{\alpha b}^{\infty }dq\,\left( \coth q-1\right) 
e^{-\frac{2qy}{b}}.
\label{N22-}
\end{eqnarray}
In the next section, we will show that 
$N_{22}^{+}(x,y,a,b)$ has surface divergences 
when $y \rightarrow b$ for all x, and also 
two edge divergences, one at $x=0,$ 
$
y=b$ and the other at $x=a$, $y=b$. We note also that 
$N_{22}^{-}(x,y,a,b)$
is finite everywhere. Using a combination of dimensional and zeta 
function regularization, in the next section we will present the general 
method to regularize the one-loop   
vacuum field fluctuations in the rectangular waveguide.

\section{Analysis of the surface divergences in the one-loop 
vacuum fluctuations.}

The purpose of this section is to present a general method to 
analytic regularize the one-loop vacuum fluctuations associated 
with the confined scalar field. We first present the 
structure of the surface and edge divergences of the 
vacuum fluctuations associated with a massless scalar field in 
a four dimensional Euclidean space.
As we discussed in the previous section, it is possible to write the one-loop 
vacuum fluctuations as
%
\begin{equation}
T_{DD}(x,y,a,b,d)=T(a,b,d)+T(x,a,b,d)+T(y,a,b,d)+T(x,y,a,b,d)
\end{equation}

The first expression that we have to deal with is $T(a,b,d)$. As we discussed 
in the previous section, the analytic structure of $T(a,b,d)$ was 
carefully analysed by Kirsten \cite{kirst} and it is not necessary to 
repeat the calculation again. The second term of the above equation and 
third one can
be written respectively in terms of $R_{1}(a,b,d)$, $R_{2}(x,a,b,d)$, $
I_{1}(a,b,d)$ and finally $I_{2}(y,a,b,d)$. The polar 
structure of $R_{1}(a,b,d)$ and 
$I_{1}(a,b,d)$ can be found in the literature, 
and we will not repeat the analysis that was done 
in these papers. The next quantity that we have to regularize is 
$R_{2}(x,a,b,d)$
and also $I_{2}(y,a,b,d)$. Since both cases are equivalent, let us study 
only the expression given by $R_{2}(x,a,b,d)$.
It is instructive to study first the simpler 
case $b\gg a$ which  will give us an 
indication of the behavior that may be encounter. Let us 
now proceed with the calculations in a general d-dimensional 
Euclidean space.
%
%
Defining $
R_{2}(x,a,b,d)|_{b\gg a}=r_{2}(x,a,d)$, we have :
\begin{equation}
r_{2}(x,a,d)=\frac{1}{2\left( 2\pi \right) ^{d-1}}\int d^{d-1}p
\frac{1}{\sqrt{\vec{p}^{\,2}+m^{2}}}\frac{\cosh ( (a-2x)\sqrt{
\vec{p}^{\,2}+m^{2} )}}{\sinh
(a\sqrt{\vec{p}^{\,2}+m^{2})}}.  \label{r2}
\end{equation}
We will again use the fact that $d^{d-1}p=p^{d-2}dp\,d\Omega_{d-1}$ and $
\int \, d\Omega_{d-1}=\frac{2\pi^{\frac{d-1}{2}}}{\Gamma(\frac{d-1}{2})}$.
Let us choose now $m=0$. Defining $h_{2}(d)$ by :
\begin{equation}
h_{2}(d)=\frac{1}{2^{d-2}\pi^{\frac{d-1}{2}} \Gamma(\frac{d-1}{2})},  
\label{h2}
\end{equation}
it is possible to write $r_{2}(x,a,d)|_{m=0}$ as 
%
%
%
\begin{eqnarray}
r_{2}(x,a,d)|_{m=0} &=&\frac{1}{2}h_{2}(d)\left[\int_{0}^{\infty} dk\,
k^{d-3}(\coth ka-1)\cosh 2kx\right.  \nonumber \\
&+& \left.\int_{0}^{\infty} dk\, k^{d-3}e^{-2kx}\right].
\label{ult11}
\end{eqnarray}
In the first integral for large $k$, $(\coth ka-1)$ has the behavior: $
(\coth ka-1) \sim e^{-2ka}$. Moreover, the second integral in the above
equation is ultraviolet finite for $x\neq 0$. Let us define $t=ka$ and 
$q=kx$
in the first and second integrals above respectively. Then 
Eq.(\ref{ult11})
becomes :
\begin{eqnarray}
r_{2}(x,a,d)|_{m=0} &=&\frac{1}{2a^{d-2}}h_{2}(d)\int_{0}^{\infty}
dt \,t^{d-3}(\coth t-1) \cosh(\frac{2xt}{a})  \nonumber \\
&+&\frac{1}{2x^{d-2}} h_{2}(d)\int_{0}^{\infty} dq \,q^{d-3}e^{-2q}.
 \label{ult2}
\end{eqnarray}
The second term in the above equation gives us the well known result that
for a massless scalar field 
$\left<\varphi^{2}(x)\right>$
diverges as $
\frac{1}{x^{2}}$ (in a four-dimensional space) as we approach 
the plate \cite{21}.  
In order to analyze the behavior of  $r_{2}(x,a,d)|_{m=0}$
around $x=a$, let us use the following integral representation 
of the gamma function,
\begin{equation}
\int_{0}^{\infty} dt \,t^{\mu-1}e^{-\nu
t}=\frac1{\nu^{\mu}}\Gamma(\mu), \,\,\,\,Re(\mu)>0,\,\,\,Re(\nu)>0
\label{J1}
\end{equation}
and also the following integral representation
of the product of the Gamma function times the Hurwitz zeta function 
\begin{equation}
\int_{0}^{\infty} dt \,t^{\mu-1}e^{-\alpha t}(\coth
t-1)=2^{1-\mu}\Gamma(\mu) 
\zeta(\mu,\frac{\alpha}{2}+1)\,\,\,\,Re(\alpha)>0,\,
\,\,Re(\mu)>1,  \label{J2}
\end{equation}
where $\zeta(s,u)$ is the Hurwitz zeta function defined by \cite{grad} 
\begin{equation}
\zeta(s,u)=\sum_{n=0}^{\infty}\frac{1}{(n+u)^{s}},\,\,\,\,Re(s)>1,
\,\,\,\,\, u \neq 0,-1,-2...  \label{na}
\end{equation}
Then, using Eqs.(\ref{J1}), (\ref{J2}) and (\ref{na}) in Eq.(\ref{ult2})
we have that :
\begin{eqnarray}
r_{2}(x,a,d)|_{m=0}&=&\frac{1}{2}h_{2}(d)\frac{1}{a^{d-2}} \left[2^{2-d}
\Gamma(d-2)\left(\zeta(d-2,\frac{x}{a}+1)+ \zeta(d-2,-\frac{x}{a}
+1)\right)\right]  \nonumber \\
&+& \frac{1}{(2x)^{d-2}}h_{2}(d)\Gamma(d-2).  \label{fim}
\end{eqnarray}
Using the definition of the Hurwitz zeta function, it is evident that : 
\begin{eqnarray}
& &\frac{1}{a^{d-2}} \left(\zeta(d-2,\frac{x}{a}+1)+ 
\zeta(d-2,-\frac{x}{a}
+1)\right)=  \nonumber \\
& &\frac{1}{a^{d-2}}\sum_{n=0}^{\infty} \frac{1}{\left(n+(1+\frac{x}{a}
)\right)^{d-2}}+\frac1{(a-x)^{d-2}}+ \frac{1}{a^{d-2}}\sum_{n=1}^{\infty} 
\frac{1}{\left(n+(1-\frac{x}{a})\right)^{d-2}}.  \label{fim2}
\end{eqnarray}
We see that the regularized $r_{2}(x,a,d)|_{m=0}$ has two poles of order $
(d-2) $ in $x=0$ and in $x=a$. Note that the residues of the poles in 
$x=0$
and in $x=a$ are $a$-independent. The same analysis can be done for 
$I_{2}(y,a,b,d)$ assuming $a\gg b$. In the next section the general case 
will be studied. 

Let us finally analyze $N_{21}(x,y,a)$ and $N_{22}(x,y,a,b)$, given 
respectively by 
\begin{equation}
N_{21}(x,y,a)=-\frac{1}{8ay}\sum_{n=1}^{\infty }\cos (\frac{2n\pi 
x}{a})e^{-
\frac{2n\pi y}{a}}
\label{bam}
\end{equation} 
and
\begin{equation}
N_{22}(x,y,a,b)=\frac{1}{4ab}\sum_{n=1}^{\infty }\cos (\frac{2n\pi x}{a}
)\int_{\frac{n\pi }{a}b}^{\infty }dq\left( \coth q-1\right) 
\cosh (\frac{2qy}{b}).
\end{equation}
To find the analytic structure of $N_{21}(x,y,a)$ we can expand the 
general
term in the sum in power series, commute the two summations, and use
analytic continuation in the zeta function that will appear. The process
will in general produce an extra term, which is generated by commuting the 
convergent
exponential summation $\sum_{m}$ with the new divergent summation $
\sum_{k}$ (for details see for e.g. refs. \cite{actor2} \cite{weldon} \cite
{elizalde}). In our case this term vanishes due to the power of $n$.
Let us express the sum that appears in Eq.(\ref{bam}) in terms of the 
complex variable $z=ix-y$: 
\begin{equation}
N_{21}(x,y,a)=-\frac{1}{8ay}Re\left\{ \sum_{n=1}^{\infty }\exp
( \frac{2n\pi z}{a}) \right\}.
\end{equation}
Expanding around $z=0$, (using the Bernoulli expansion) will produce : 
\begin{equation}
N_{21}(x,y,a)=-\frac{1}{8ay}Re\left\{-\frac{a}{2\pi z}+\frac{1}{2}+
\sum_{k=2}^{\infty}\frac{B_{k}}{k!}(-\frac{2\pi z}{a})^{k}\right\}.
\end{equation} 
%
%
We see that the edge divergence appears in the term $1/z$. Taking the real
part : 
\begin{equation}
N_{21}(x,y,a)=-\frac{1}{16\pi \left(y^{2}+x^{2} \right)}-\frac{1}{8ay}
Re\left\{ f_{1}(z)\right\},
\end{equation}
where $f_{1}(z)$ is an entire function of $z$ and is given by : 
\begin{equation}
f_{1}(z)=Re\left\{
\frac{1}{2}+
\sum_{k=2}^{\infty}\frac{B_{k}}{k!}\left(-\frac{2\pi z}{a}\right)^{k}
\right\}.
\end{equation}
Expanding around $z=ia$ will produce : 
\begin{equation}
N_{21}(x,y,a)=-\frac{1}{8ay}Re\left\{ \sum_{k=0}^{\infty }\left( 
\frac{2\pi }{a}\right) ^{k}\frac{(z-ia)^{k}}{k!}\zeta (-k)-\left( 
\frac{a}{2\pi }\right) (z-ia)^{-1}\right\}.
\end{equation}
Taking the real part of ${1/(}{z-ia)}$, we have : 
\begin{equation}
N_{21}(x,y,a)=-\frac{1}{16\pi \left( y^{2}+(x-a)^{2} \right)}-\frac{1}{8ay} 
f_{2}(z),
\end{equation}
where $f_{2}(z)$ is also an entire function of $z$,
and is given by: 
\begin{equation}
f_{2}(z)=Re\left\{ \sum_{k=0}^{\infty }\left( \frac{2\pi }{a}\right) ^{k}
\frac{(z-ia)^{k}}{k!}\zeta (-k)\right\}.
\end{equation}
To find the analytic structure of $N_{22}(x,y,a,b)$, it is enough to 
analyse
the quantity $N_{22}^{+}(x,y,a,b)$, which is given by the Eq.(\ref{N22+})
(the other quantity $N_{22}^{-}(x,y,a,b)$ is finite everywhere). To
calculate the integral in Eq.(\ref{N22+}), we can express $\coth (bv)$
in exponential functions and expand the integrand in power series. The
integral can be easily evaluated, and the result is given by 
\begin{equation}
N_{22}^{+}(x,y,a,b)=-\frac{1}{8ab}\sum_{n=1}^{\infty }\cos (\frac{2n\pi 
x}{a}
)\sum_{k=0}^{\infty }\frac{e^{-2\left( 1-\frac{y}{b}+k\right) \frac{n\pi 
}{a}
b}}{\left( 1-\frac{y}{b}+k\right)}.
\end{equation}
This result can be written in terms of the Lerch's trancendent function  
$\Phi(z,c,v)$ which is 
defined as follows  \cite{lech} :
\begin{equation}
\Phi(z,c,v) =\sum_{k=0}^{\infty}
\frac{z^{k}}{\left(v+k\right)^{c}}.
\end{equation}
which is valid for $|z|<1$. By analytic continuation it can be extended to
the whole complex plane. This function has 
singularities at $z=1$ and $c=0$ or $
c=1$ and when $v$ is a non-positive integer and Re$(c)$ is also 
non-positive.
We note that only the term $k=0$ has a surface divergence at $y=b$. The
remaining part of the series in $k$ is finite everywhere and we call it $
F(x,y,a,b)$. Using the same procedure used to analyse $N_{21}(x,y,a),$ we
can define $z=(b-y)+ix$ and expand the term $k=0$ around $z=0$ or around $
z=ia$. Therefore we have 
\begin{equation}
N_{22}^{+}(x,y,a,b)=B(x,y,a,b)+F(x,y,a,b),
\end{equation}
where the first expansion $B(x,y,a,b)$ is given by 
\begin{equation}
B(x,y,a,b)=-\frac{1}{16\pi \left( x^{2}+(y-b)^{2} \right)}-
\frac{1}{8a(y-b)}h_{1}(z),
\end{equation}
where $h_{1}(z)$ is an entire function given by
\begin{equation}
h_{1}(z)=Re\left\{ \sum_{k=0}^{\infty }\left( \frac{2\pi }{a}\right) ^{k}
\frac{(z)^{k}}{k!}\zeta (-k)\right\}.
\end{equation}
We note that in this case $B(x,y,a,b)$ has a edge divergence 
for $x=0,$ $y=b$. For the second expansion, $B(x,y,a,b)$ is given by 
\begin{equation}
B(x,y,a,b)=-\frac{1}{16\pi \left( (x-a)^{2}+(y-b)^{2} 
\right)}-\frac{1}{8a(y-b)}h_{2}(z),
\end{equation}
where $h_{2}(z)$ is an entire function given by 
\begin{equation}
h_{2}(z)=Re\left\{ \sum_{k=0}^{\infty }\left( \frac{2\pi }{a}\right) ^{k}
\frac{(z-ia)^{k}}{k!}\zeta (-k)\right\}.
\end{equation}
We note that in this case $B(x,y,a,b)$ has a edge divergence for $x=a,$ $
y=b$. A general picture that emerges  
from the above discussion is the following: 
we have found that  in order to eliminate the 
ultraviolet divergences of the one-loop vacuum fluctuations 
we have to introduce not only the 
usual bulk counterterms, but also
counterterms concentrated on the boundaries. 
In the next section we will show how it is possible to 
generalize our local analysis 
for a massive field defined in a higher dimensional Euclidean space.

\section{The one-loop vacuum fluctuation in the waveguide in a 
higher dimensional Euclidean space.}

In this section we will discuss how to generalize  
our calculations to a higher dimensional Euclidean space. 
We are interested 
in investigating the one-loop vacuum fluctuations associated with a 
massive scalar field defined in a generic d-dimensional Euclidean space, 
where $d-2$ coordinates 
are unbounded while the first two lie in some finite interval. 
As discussed in the previous sections, we 
found that the vacuum fluctuations can be expressed as :
\begin{eqnarray}
\left<\varphi^{2}(x,y)\right>&=&\frac{1}{4}T(a,b,d)+
\frac{1}{2}(R_{1}(a,b,d)+I_{1}(a,b,d))+\nonumber\\
&&
\frac{1}{2}\left(R_{2}(x,a,b,d)+I_{2}(y,a,b,d)\right)+N(x,y,a,b,d).
\label{aa1}
\end{eqnarray}
In the 
previous sections, we studied the expression given by $T(a,b,d)$,
$R_{1}(a,b,d)$ and also
$I_{1}(a,b,d)$ in the massless case, which
demands bulk counterterms.
In the following, we will present a detailed analysis of 
the contributions of the vacuum fluctuations 
that demand counterterms concentrated on the 
boundaries. We will show that some of  the terms of the one-loop vacuum 
fluctuations, can 
be expressed in terms of Bessel functions and also Hurwitz zeta functions.

From the previous analysis for the massless field in a 
four-dimensional Euclidean space, we found that to perform our calculations
in a generic d-dimensional space, for the massive field, 
it is natural to write  $I_{2}(y,a,b,d)$ in the following way :  
\begin{equation}
I_{2}(y,a,b,d)=I_{21}(y,a,d)+I_{22}(y,a,b,d), 
\label{aa2}
\end{equation}
where the functions $I_{21}(y,a,d)$ and $I_{22}(y,a,b,d)$ 
are given by :
\begin{equation}
I_{21}(y,a,d)=\frac{1}{a}h(d)\sum_{n=1}^{\infty} 
\int_{\alpha }^{\infty} 
dv(v^{2}-m^{2}-(\frac{n\pi}{a})^{2})^{\frac{d-4}{2}}
e^{-2vy},
\label{aa3}
\end{equation}
and
\begin{equation}
I_{22}(y,a,b,d)=
\frac{1}{a}h(d)\sum_{n=1}^{\infty} 
\int_{\alpha }^{\infty} dv(v^{2}-m^{2}-
(\frac{n\pi}{a})^{2})^{\frac{d-4}{2}}
(\coth bv-1)\cosh 2vy.
\label{aa4}
\end{equation}
The lower limit of both integrals is given by 
$\alpha$ defined in Eq.(\ref{alfa}) and 
the coefficient $h(d)$ is an entire function defined in Eq.(\ref{hd}). 
Using the following integral 
representation of the 
modified Bessel functions of third kind, or the Macdonald's 
functions $K_{\nu}(x)$ :
\begin{equation}
\int_{u}^{\infty}(x^{2}-u^{2})^{\nu-1}e^{-\mu x} \,dx=
\frac{1}{\sqrt{\pi}}(\frac{2u}{\mu})^{\nu-\frac{1}{2}}\Gamma(\nu)
K_{\nu-\frac{1}{2}}(u\mu),
\label{aa5}
\end{equation}
which is valid for for $u>0$, $Re\,\mu>0$ and $Re\,\nu>0$, 
$I_{21}(y,a,b,d)$ 
can be written in terms of these functions. A simple substitution gives
\begin{equation}
I_{21}(y,a,b,d)=\frac{1}{a}\frac{1}{(2\sqrt{\pi})^{d-1}}
\sum_{n=1}^{\infty}
(\frac{\alpha}{y})^{\frac{d-3}{2}}
K_{\frac{d-3}{2}}(2\alpha y).
\label{aa6}
\end{equation}
Equation (\ref{aa6}) can 
not be evaluated exactly, even for the massless case,
but for small arguments of the Bessel
function it is possible to show how does $I_{21}(y,a,d)$ contributes to 
the surface divergences of the one-loop vacuum fluctuations.

The expression $I_{22}(y,a,b,d)$  
can also be computed in the both cases, the 
massive and the massless one. To ilustrate the method that we 
discussed in previous sections,
let us present the massless case calculation. The massive case 
follows the same procedure. 
To calculate $I_{22}(y,a,b,d)|_{m=0}$ note that after change  
of variables $I_{22}(y,a,b,d)|_{m=0}$, 
contains a power of a binomial. When $d$ is even (for $d>4$)
the power is an integer and the use of the Newton's binomial theorem will 
give a very direct way to generalize our further results. When $d$ is 
odd $(d>4)$, the expansion on the binomial yields an infinite power 
series. The same technique was used in refs. 
\cite{ss} \cite{claude} \cite{nucle}. 
It is worthwhile to remark that the study of 
even case follows the same procedure, only with a finite numbers of 
terms in the binomial expansion. Consequently, let us study the
most interesting case, i.e the odd dimensional case.
Let us use the following series representation for 
$(1-(\frac{n\pi}{va})^{2})^{\frac{d-4}{2}}$ :
\begin{equation}
(1-(\frac{n\pi}{va})^{2})^{\frac{d-4}{2}}=
\sum_{k=0}^{\infty}(-1)^{k}C^{k}_{p}(\frac{n\pi}{va})^{2k},
\label{aa7}
\end{equation}
where the $C^{k}_{p}$ are the generalizations of the binomial coeficients,
given by: $C_{p}^{0}=1$,$\,C_{p}^{1}=
\frac{p}{1!}$,$\,C_{p}^{2}=\frac{1}{2!}p(p-1)$,...
$\,C_{p}^{k}=\frac{1}{k!}(p(p-1)..(p-k+1))$ and $p=\frac{d-4}{2}$. 
Note that the generalization of the binomial series is valid even 
for any complex exponent $p$. In other words, for $v>\frac{n\pi}{a}$ we 
have an everywhere 
convergent power series in $p$, hence a continuous function on 
$p$ in the complex $d$ plane.
In the following, 
it is convenient to define the quantities: $C^{(1)}(d,k)=
(-1)^{k}C^{k}_{p}h(d)$, $C^{(2)}(d,k)=\pi^{2k}C^{(1)}(d,k)$ and 
finally $C^{(3)}(d,k)=\frac{\Gamma(d-3-2k)}{2^{d-3-2k}} C^{(2)}(d,k)$.
After a change of variables, using Eq.(\ref{aa7}) in Eq.(\ref{aa4}), 
$I_{22}(y;a,b,d)|_{m=0}$ becomes
\begin{equation}
I_{22}(y,a,b,d)|_{m=0}=
\frac{1}{ab^{d-3}}\sum_{k=0}^{\infty}C^{(2)}(d,k)(\frac{b}{a})^{2k}
\sum_{n=1}^{\infty}
n^{2k}\int_{\frac{n\pi b}{a}}^{\infty}du\,
u^{d-4-2k}(\coth u-1)\cosh (\frac{2uy}{b}),
\label{007}
\end{equation}
where $C^{(2)}(d,k)$ is an entire function in the complex $d$ plane.
The integral that appear in Eq. (\ref{007})
cannot be evaluated explicity in terms of well 
known functions. Nevertheless it is possible to write Eq.(\ref{007}) 
in a convenient way where the structure of the divergences near the plate 
when $y\rightarrow b$ appear. Let us split the $I_{22}(y,a,b,d)|_{m=0}$ in the 
following way :
\begin{equation}
I_{22}(y,a,b,d)|_{m=0}=I_{22}^{<}(y,a,b,d)+I_{22}^{>}(y,a,b,d),
\label{i22}
\end{equation}
where
\begin{equation}
I_{22}^{<}(y,a,b,d)=\frac{1}{ab^{d-3}}
\sum_{k=0}^{k < \frac{d-4}{2}}
C^{(2)}(d,k)(\frac{b}{a})^{2k}\sum_{n=1}^{\infty}n^{2k}
\int_{\frac{n\pi b}{a}}^{\infty} du\, u^{d-4-2k}
(\coth u -1)\cosh(\frac{2uy}{b}), 
\label{i23}
\end{equation}
and 
\begin{equation}
I_{22}^{>}(y,a,b,d)=\frac{1}{ab^{d-3}}
\sum_{k\geq \frac{d-4}{2}}^{\infty}C^{(2)}(d,k)
(\frac{b}{a})^{2k}\sum_{n=1}^{\infty}n^{2k}
\int_{\frac{n\pi b}{a}}^{\infty} du\, u^{d-4-2k}
(\coth u -1)\cosh(\frac{2uy}{b}). 
\label{i24}
\end{equation}
Although the integral that appears in Eq.(\ref{i24}) can not be 
expressed in tems of known functions, 
the leading divergences of $I_{22}(y,a,b,d)|_{m=0}$ are contained in
the Eq.(\ref{i23}).
Consequently, let us analyze the divergences that appear in Eq.(\ref{i23}). 
Using the fact that  
$\int^{\infty}_{\alpha}f(u)du=\int_{0}^
{\infty}f(u)du-\int_{0}^{\alpha}f(u)du$,
we reexpress Eq.(\ref{i23}) as
\begin{equation}
I_{22}^{<}(y,a,b,d)=I_{22}^{a}(y,a,b,d)+I_{22}^{b}(y,a,b,d),
\label{i255}
\end{equation}
where
\begin{equation}
I_{22}^{a}(y,a,b,d)=\frac{1}{ab^{d-3}}
\sum^{k< \frac{d-4}{2}}_{k=0}
C^{(2)}(d,k)(\frac{b}{a})^{2k}\sum^{\infty}_{n=1}n^{2k}
\int_{0}^{\infty} du\, u^{d-4-2k}
(\coth u -1)\cosh(\frac{2uy}{b}), 
\label{i233}
\end{equation}
and 
\begin{equation}
I_{22}^{b}(y,a,b,d)=-\frac{1}{ab^{d-3}}
\sum^{k< \frac{d-4}{2}}_{k=0}
C^{(2)}(d,k)(\frac{b}{a})^{2k}\sum_{n=1}^{\infty}n^{2k}
\int_{0}^{\frac{n\pi b}{a}} du\, u^{d-4-2k}
(\coth u -1)\cosh(\frac{2uy}{b}). 
\label{i244}
\end{equation}
Expanding $cosh (\frac{2uy}{b})$ in powers series, 
the integral that appears in $I_{22}^{b}(y,a,b,d)$ is 
a Debye integral. Although this can be evaluated, 
the study of $I_{22}^{a}(y,a,b,d)$ is sufficient to 
obtain the divergences of $I_{22}^{<}(y,a,b,d)$. Again, we note that  
the integral that 
appear in Eq.(\ref{i233}) can be written 
in tems of products of zeta, Gamma function 
and Hurwitz zeta function. Thus we have :  
\begin{eqnarray}
I_{22}^{<}(y,a,b,d)|_{m=0}&=&
-\frac{1}{ab^{d-3}}
\sum^{k<\frac{d-4}{2}}_{k=0}
C^{(2)}(d,k)(\frac{b}{a})^{2k}\sum_{n=1}^{\infty}n^{2k}
\int_{0}^{\frac{n\pi b}{a}} du\, u^{d-4-2k}
(\coth u -1)\cosh(\frac{2uy}{b})
\nonumber\\
&&+\frac{1}{2ab^{d-3}}
C^{(3)}(d,0)
\left(\zeta(d-3,-\frac{y}{b}+1)+
\zeta(d-3,\frac{y}{b}+1)\right).
\label{nnn}
\end{eqnarray}
Since $C^{(3)}(d,0)$ is an entire function in the complex $d$ plane, 
it is clear that we have the same surface divergences that we 
studied before. To study the massive case $I_{22}(y,a,b,d)$
we have to use that same procedures. 
Let us show how our  procedure can be systematically applied 
in the other expressions.
From the previous calculations we 
express the contribution to the vacuum fluctuations 
that contains edge 
divergences giving by $N(x,y,a,b,d)$ in the following way :
\begin{equation}
N(x,y,a,b,d)=N_{11}(a,b,d)+N_{12}(x,a,b,d)+N_{2}(x,y,a,b,d),
\label{aa9}
\end{equation}
where the $N_{11}(a,b,d)$ term contains contributions to  
the vacuum fluctuations that demands bulk counterterms
in order to renormalize the one-loop vacuum fluctuations.
The next contribution that we have to 
study is $N_{12}(x,a,b,d)$. Starting from Eq.(\ref{n12})
integrating over the solid angle and changing variables, it is also convenient 
to split $N_{12}(x,a,b,d)$ in the following way :
%
%
%
\begin{equation}
N_{12}(x,a,b,d)= 
N_{12}^{a}(x,b,d)+N_{12}^{b}(x,a,b,d),
\label{aa11}
\end{equation}
where each of the terms of Eq.(\ref{aa11}) are given respectively by
\begin{equation}
N_{12}^{a}(x,b,d)=
-\frac{1}{4b}h(d)\int_{m}^{\infty} dv\,(v^{2}-m^{2})^{\frac{d-4}{2}}
e^{-2vx},
\label{aa12}
\end{equation}
and
\begin{equation}
N_{12}^{b}(x,a,b,d)=
-\frac{1}{4b}h(d)\int_{m}^{\infty} dv\,(v^{2}-m^{2})^{\frac{d-4}{2}}
(\coth av -1)\cosh 2vx. 
\label{aa13}
\end{equation}
We can again use the integral 
representation of the modified Bessel functions of the third kind 
given by Eq.(\ref{aa5}), to write Eq.(\ref{aa12}) as 
\begin{equation}
N_{12}^{a}(x,b,d)=-\frac{1}{b}\frac{1}{(2\sqrt{\pi})^
{d-1}}(\frac{m}{x})^{\frac{d-3}{2}}
K_{\frac{d-3}{2}}(2mx).
\label{aa14}
\end{equation}
To calculate  $N_{12}^{b}(x,a,b,d)$, let 
us study first the massless case. In this case,
first changing variables and using Eq.(\ref{J2}) we get  
\begin{equation}
N_{12}^{b}(x,a,b,d)|_{m=0}=-
\frac{1}{a^{d-3}b}\frac{h(d)}{2^{d}}
\Gamma(d-3)\left(
\zeta(d-3,\frac{x}{a}+1)+\zeta(d-3,-\frac{x}{a}+1)\right).
\label{aa15}
\end{equation}
Next we consider the massive case. Note that 
we have exactly the same situation discussed before.
After a change of variables, $N_{12}^{b}(x,a,b,d)$ 
also contains a power of a binomial. 
Let us discuss again the most interesting situation, the   
odd dimensional case. First, let us the 
same power series expansion that we used before for the quantity 
$(1-\frac{m^{2}}{v^{2}})^{\frac{d-4}{2}}$ given by
\begin{equation}
(1-\frac{m^{2}}{v^{2}})^{\frac{d-4}{2}}=
\sum_{k=0}^{\infty}(-1)^{k}C^{k}_{p}(\frac{m}{v})^{2k}.
\label{aa16}
\end{equation}
Substituting Eq.(\ref{aa16}) in Eq.(\ref{aa13}) and 
again changing variable, we have
\begin{equation}
N_{12}^{b}(x,a,b,d)=-\frac{1}{4a^{d-3}b}h(d)
\sum_{k=0}^{\infty}(-1)^{k}C^{k}_{p}(ma)^{2k}
\int_{ma}^{\infty} du\, u^{d-4-2k}
(\coth u -1)\cosh \frac{2ux}{a}. 
\label{aa17}
\end{equation}
A natural way to obtain $N_{12}^{b}(x,a,b,d)$ expressed in terms of the 
Hurwitz zeta function is 
the following. Let us split $N_{12}^{b}(x,a,b,d)$ in the following way :
\begin{equation}
N_{12}^{b}(x,a,b,d)=N_{12}^{b<}(x,a,b,d)+N_{12}^{b>}(x,a,b,d),
\label{aa18}
\end{equation}
where
\begin{equation}
N_{12}^{b<}(x,a,b,d)=-\frac{1}{4a^{d-3}b}
\sum_{k=0}^{k<\frac{d-4}{2}}C^{(1)}(d,k)(am)^{2k}
\int_{am}^{\infty} du\, u^{d-4-2k}
(\coth u -1)\cosh(\frac{2ux}{a}),
\label{aa19}
\end{equation}
and 
\begin{equation}
N_{12}^{b>}(x,a,b,d)=-\frac{1}{4a^{d-3}b}
\sum_{k\geq \frac{d-4}{2}}^{\infty}C^{(1)}(d,k)(am)^{2k}
\int_{am}^{\infty} du\, u^{d-4-2k}
(\coth u -1)\cosh(\frac{2ux}{a}). 
\label{aa20}
\end{equation}
We have the same situation that we studied before. Let us investigate first
the $N_{12}^{b<}(x,a,b,d)$. Using that $\int^{\infty}_{\alpha}f(u)du=
\int_{0}^{\infty}f(u)du-\int_{0}^{\alpha}f(u)du$, it is 
possible to write Eq.(\ref{aa19}) in the following way:
\begin{eqnarray}
N_{12}^{b<}(x,a,b,d)&=&-\frac{1}{4a^{d-3}b}
\sum_{k=0}^{k<\frac{d-4}{2}}C^{(3)}(d,k)(\frac{am}{\pi})^{2k}
\left(\zeta(d-3-2k,-\frac{x}{a}+1)+
\zeta(d-3-2k,\frac{x}{a}+1)\right)\nonumber\\
&&
+\frac{1}{4a^{d-3}b}
\sum_{k=0}^{k<\frac{d-4}{2}}
C^{(1)}(d,k)(am)^{2k}
\int_{0}^{am} du\, u^{d-4-2k}
(\coth u -1)\cosh(\frac{2ux}{a}),
\label{aa21}
\end{eqnarray}
where the singularities of
$N_{12}^{b<}(x,a,b,d)$ appear at $x\rightarrow a$.
We have to analyze the $N_{12}^{b>}(x,a,b,d)$. It is clear that in a even 
dimensional Euclidean space for $x<a$ the integral that appear in Eq.(\ref{aa20}) 
is convergent. The odd dimensional case also can be studied. In this 
case we have to expand $cosh (\frac{2uy}{a})$ in power series and it is clear that 
in the new integral that appears in Eq.(\ref{aa20}) demands a generalization 
of the Debye integrals. 
The calculations that we presented  can also be used to write
$N_{2}(x,y,a,b,d)$ in the same form that we obtained for the other expressions. 
Unfortunatelly one cannot perform the summation 
in $n$. For completeness, we will only write the 
remaining expressions. Let us start from :
\begin{equation}
N_{2}(x,y,a,b)=N_{21}(x,y,a,d)+N_{22}(x,y,a,b,d),
\label{aa22}
\end{equation}
where each term of the above equation is given by
\begin{equation}
N_{21}(x,y,a,d)=\frac{1}{2\pi a}h(d)\sum_{n=1}^{\infty}\cos(\frac{2n\pi x}{a}) 
\int_{\alpha }^{\infty} 
dv(v^{2}-m^{2}-(\frac{n\pi}{a})^{2})^{\frac{d-4}{2}}
e^{-2vy},
\label{aa23}
\end{equation}
and
\begin{eqnarray}
N_{22}(x,y,a,b,d)&=&
\frac{1}{2\pi a}h(d)\sum_{n=1}^{\infty}\cos(\frac{2n\pi x}{a})
\nonumber\\ 
&&\int_{\alpha }^{\infty} dv(v^{2}-m^{2}-
(\frac{n\pi}{a})^{2})^{\frac{d-4}{2}}
(\coth bv-1)\cosh 2vy.
\label{aa24}
\end{eqnarray}
Note that the lower limit of both integrals are given by 
$\alpha$ defined in Eq.(\ref{alfa}) and 
$h(d)$ was defined in Eq.(\ref{hd}). Let us present the 
$N_{21}(x,y,a,b,d)$ in the massless case. Using the same integral 
representation of the Bessel functions that we used before gives
\begin{equation}
N_{21}(x,y,a,d)|_{m=0}=\frac{2}{a(2\sqrt{\pi})^{d+1}}
\sum_{n=1}^{\infty}\cos(\frac{2n\pi x}{a})
\left(\frac{n\pi}{ay}\right)^{\frac{d-3}{2}}
K_{\frac{d-3}{2}}(\frac{2n\pi y}{a}).
\label{aa25}
\end{equation}
The massive case can also be presented 
but in both cases it is not possible to perform the summation in $n$.
Finally, let us present the expression for  
$N_{22}(x,y,a,b,d)$ in the massless case and for the odd dimensional case :
\begin{eqnarray}
N_{22}(x,y,a,b,d)|_{m=0}&=&
\frac{1}{2\pi ab^{d-3}}\sum_{k=0}^{\infty}C^{(2)}(d,k)(\frac{b}{a})^{2k}
\sum_{n=1}^{\infty}n^{2k}\cos(\frac{2n\pi x}{a})
\nonumber\\
&&\int_{\frac{n\pi b}{a}}^{\infty}du\,
u^{d-4-2k}(\coth u-1)\cosh (\frac{2uy}{b}).
\label{aa26}
\end{eqnarray}
The same procedure that we used in $I_{22}(y,a,b,d)$,  
can be repeated again. Unfortunatelly one can not perform the summation 
in $n$ that appears in Eq.(\ref{aa26}). Nevertheless the same analysis 
that we did in the end of section III can be performed to study the 
edge divergences that appear in the one-loop vacuum fluctuations associated 
with the scalar field confined in the interior of a waveguide in a higher 
dimesional Euclidean space.

\section{Discussions and conclusions}

In this paper, we first obtained the regularized 
one-loop vacuum fluctuations 
associated with a massless scalar field defined in the interior of 
an infinity  waveguide of rectangular cross section 
in a four-dimensional Euclidean space. Then, 
we discussed how it is possible to generalize our results for a massive
field defined in a higher dimensional Euclidean space.

Let us summarize our motivations and the results.
First, in the rectangular waveguide configuration the calculations can be done 
analytically, and since in this geometric 
configuration the electromagnetic 
field can be described by two 
massless scalar fields with Dirichlet and Neumann boundary 
conditions, our calculations can be used to describe the vacuum fluctuations 
associated with the electromagnetic field in the interior of the 
waveguide. Also, the rectangular geometries are very 
convenient to carry out experiments 
on the effect of boundaries on the atomic levels of atoms and  
on the rate of spontaneous emission. 
It is well known that in waveguides, it is possible to obtain situations 
where the spontaneous emission of atoms can be suppresed and also 
enhanced. Finally, the calculation of the regularized one-loop vacuum 
fluctuations in the infinite 
waveguide of rectangular cross section using the combination of dimensional 
and zeta function analytic regularization has not been discussed 
in the literature, at least as far as we know.

We first rederive a well-know result that 
surface and edge divergences 
appear in the one-loop vacuum fluctuations  
as consequence of the uncertaintly principle. There are
at least two different possible solutions that can eliminate these 
divergences. The first is to take into account that real materials have 
imperfect conductivity at high frequencies. 
As was stressed by many authors, the infinities that appear in renormalized 
values of local observables for the ideal conductor (or perfect mirror)
represent a breakdown of the perfect-conductor approximation. A wavelength 
cutoff corresponding to the finite plasma frequency must be included.  
The 
second would be given by a quantum mechanical treatment of the 
boundary conditions \cite{ls}. It was shown \cite{ls}
that position fluctuations of a 
reflecting boundary also remove divergences in the renormalized values 
of local observables at least in the 
flat plate configuration. A logical going is to  
use an analytic regularization procedure
to identify these divergent terms. As we discussed 
counterterms concentrated on the 
boundaries  produce a finite one-loop vacuum fluctuations in the 
interior of the waveguide.   
A natural extension of this paper is to 
go beyond the one-loop approximation,
investigating interacting field 
models in the presence of a waveguide in a 
d-dimensional Euclidean space. 
This topic is  under investigation by the authors.

\section{Acknowledgement}

We would like to thank B.Schroer and L.H.Ford for several helpful
discussions. N.F.Svaiter would like to acknowledge the hospitality of the 
Center of Theoretical Physics, Laboratory for 
Nuclear Science and Department of Physics of the  
Massachusetts Institute of Technology,
where part of this work was carried out. This 
paper was supported in part by Conselho Nacional de
Desenvolvimento Cientifico e Tecnologico do Brazil 
(CNPq) and also by funds provided by the 
U.S.Department of Energy(D.O.E) under cooperative 
research agreement DF-FC02-94ER40810.

\begin{appendix}
\makeatletter
\@addtoreset{equation}{section}
\makeatother
\renewcommand{\theequation}{\thesection.\arabic{equation}}

\section{The one-loop vacuum fluctuations and the 
generalized zeta function method}

In aim of this appendix is to discuss the link between the one-loop vacuum 
fluctuations and the generalized zeta function 
method \cite{haw} \cite{dow} \cite{voros},
as a standard formalism to regularize products and determinants. 
Since our prime goal in section IV was to 
calculate the one-loop vacuum fluctuations in the massive model, we will 
present the formal relation between the one-loop vacuum fluctuations and the 
effective action in the one-loop approximation. 
We are following the treatment used by Dittrich and Reuter \cite{igual}.

Let us consider the generating 
functional of the complete Schwinger 
functions for a scalar field in a d-dimensional 
Euclidean space:
\begin{equation}
Z(J)=\int D\varphi\, e^{-S[\varphi]+\int d^{d}x\, J(x)\varphi(x)},
\label{e1}
\end{equation}
where $D\varphi$ is the appropriate measure, and $S[\varphi]$ is the classical 
action associated with the scalar field. The quantity $Z(J)$ can be regarded 
as the functional integral representation for the imaginary time evolution
operator $\left<\varphi_{2}|U(t_{2},t_{1})|\varphi_{1}\right>$ with the boundary 
conditions $\varphi(t_{1},\vec{x})=\varphi_{1}(\vec{x})$ and  
$\varphi(t_{2},\vec{x})=\varphi_{2}(\vec{x})$. The 
quantity $Z(J)$ gives the transition 
amplitude from the initial state $|\varphi_{1}>$
to a final state  $|\varphi_{2}>$ in the presence of some scalar source 
of compact suport. As usual $W(J)$, the generating functional 
of the connected correlation functions shall be given by $W(J)=ln Z(J)$. In a 
free theory the partition function $Z(J)$ and also $W(J)$ can be 
calculated exactly. 
In the
presence of the waveguide we must assume that the path integral must 
be taken over the space of functions that vanish in the boundaries of the 
waveguide. Since our aim is to discus the non-interacting theory, 
let us assume that $J=0$. Thus, defining the d-dimensional laplacian by
$\Delta$, we have:
\begin{equation}
Z(J)|_{J=0}=\int_{Dirichlet} D\varphi\, e^{-\frac{1}{2}\int\,d^{d}x 
\,\varphi(x)(-\Delta+m^{2})\varphi(x)},
\label{e2}
\end{equation}
``Dirichlet'' in the path integral means that 
we are performing the path integral over functions that vanish on the 
boundaries. Thus the generating functional can be written as
\begin{equation}
Z(J)|_{J=0}=N(det(-\Delta+m^{2}))^{-\frac{1}{2}},
\label{e3}
\end{equation}
where $N$ is a normalization factor that does not contributes to the free
energy. For simplicity, let us discuss the four-dimensional case. 
The global generalized zeta function is defined by
\begin{equation}
\zeta_{-\Delta+m^{2}}(s)=N'\int dk_{0}\int dk_{1}\sum_{n,n^{\prime
}=1}^{\infty}(k_{0}^{2}+k_{1}^{2}+(\frac{n\pi }{a})^{2}+(\frac{
n^{\prime }\pi }{b})^{2}+m^{2})^{-s},
\label{e6}
\end{equation}
where $N'$ is a normalization constant.
It follows that
\begin{equation}
Z(J=0)=exp{(\frac{1}{2}\zeta^{'}_{-\Delta+m^{2}}(0))},
\label{e4}
\end{equation}
where $\zeta^{'}(0)=\frac{d}{ds}\zeta(s)|_{s=0}$ and 
the operator $(-\Delta+m^{2})$ has the spectrum
\begin{equation}
(k_{0}^{2}+k_{1}^{2}+(\frac{n\pi }{a})^{2}+(\frac{
n^{\prime }\pi }{b})^{2}+m^{2}), k_{0},k_{1}\in \Re, n,n'\in N.
\label{e5}
\end{equation}
We can use the same procedure to define the local generalized zeta function 
which is related with the effective action.
The local generalized zeta function is given by
\begin{eqnarray}
\zeta_{-\Delta+m^{2}}(s,x,y)&=& N'\int dk_{0}\int dk_{1}\sum_{n,n^{\prime
}=1}^{\infty}\frac{\sin ^{2}( \frac{n\pi x}{a})
\sin ^{2}( \frac{n^{\prime }\pi y}{
b})}{(k_{0}^{2}+k_{1}^{2}+(\frac{n\pi }{a})^{2}+(\frac{
n^{\prime }\pi }{b})^{2}+m^{2})^{s}},
\label{e7}
\end{eqnarray}
where $x$ and $y$ are two cartesian 
coordinates of one point of the infinite 
waveguide with rectangular crosss section.
Thus we have the formal expression:
\begin{equation}
(-1)^{s}(\frac{\partial}{\partial m^{2}})^{s}\left<\varphi^{2}(x,y)\right>
=\zeta(s,x,y).
\label{e8}
\end{equation}
To give a precise meaning of the derivative one can use for example
the Liouville's concept of fractional derivative \cite{ll}.
This result show that the  one-loop vacuum 
fluctuations determines the effective action in the one-loop level.

\end{appendix}

\end{document}